\documentclass[]{aastex7}
\received{May 16, 2025}
\revised{June 27, 2025}
\accepted{July 2, 2025}
\begin{document}

\title{On Dust Devil Diameters, Occurrence Rates, and Activity}

\correspondingauthor{Brian Jackson}
\email{bjackson@boisestate.edu}

\author[0000-0002-9495-9700]{Brian Jackson}
\email{bjackson@boisestate.edu}
\affiliation{Department of Physics, Boise State University, 1910 University Drive, Boise ID 83725-1570 USA}
\affiliation{Carl Sagan Center, SETI Institute, Mountain View, CA, United States}

\author[0000-0001-8116-4901]{Lori Fenton}
\email{lfenton@seti.org}
\affiliation{Carl Sagan Center, SETI Institute, Mountain View, CA, United States}

\author[0000-0001-8528-4644]{Ralph Lorenz}
\email{Ralph.Lorenz@jhuapl.edu}
\affiliation{Johns Hopkins Applied Physics Laboratory, 1100 Johns Hopkins Road, Laurel, MD, USA}

\author{Chelle Szurgot}
\email{michelleszurgot@boisestate.edu}
\affiliation{Department of Physics, Boise State University, 1910 University Drive, Boise ID 83725-1570 USA}
\affiliation{Department of Geosciences, Boise State University, 1910 University Drive, Boise ID 83725-1570 USA}

\author{Joshua Gambill}
\email{joshuagambill@boisestate.edu}
\affiliation{Department of Physics, Boise State University, 1910 University Drive, Boise ID 83725-1570 USA}

\author{Gwendolyn Arzaga}
\email{garzaga@wesleyan.edu}
\affiliation{Wesleyan University, 45 Wyllys Avenue, Middletown, CT 06459}

\begin{abstract}

\added{As a phenomenon that occurs on Earth and on Mars, t}he diameter of a dust devil helps determine the amount of dust the devil injects into the atmosphere \added{for both worlds} -- for a given dust flux \added{density (dust lifted per area per time)}, a wider devil will lift more dust into the air. However, the factors that determine a dust devil's diameter $D$ and how it might relate to ambient conditions have remained unclear. Moreover, estimating the contribution to an atmospheric dust budget from a population of dust devils with a range of diameters requires an accurate assessment of the \added{differential} diameter distribution, but considerable work has yet to reveal the best representation or explain its physical basis. In this study, we propose that this distribution follows a power-law $\propto D^{-5/3}$ and provide a simple physical explanation for why the distribution takes this form. By fitting diameter distributions of martian dust devil diameters reported in several studies, we show that the data from several studies support this proposed form. Using a previous model that treats dust devils as thermodynamic heat engines, we also show that the areal density of dust devils (number per unit area) $N_0$ scales with the product of their thermodynamic efficiency $\eta$ and the sensible heat flux $F_{\rm s}$ as $N_0 \propto \eta F_{\rm s}$.
\end{abstract}



\section{Introduction}\label{sec:Introduction}

Whirling vortices laden with aerosols, dust devils stalk the dusty surface of Mars \added{and arid regions on the Earth}. As they lift dust from the surface into the atmosphere, they contribute to the perpetual background dust that obscures the martian horizon. Not only does this ubiquitous haze impact operation of missions on Mars \citep[e.g.,][]{2023LPICo2806.2467G}, it also absorbs sunlight and heats the atmosphere \citep{2017acm..book..229K}. This atmospheric warming can \added{reduce the effectiveness of} the atmospheric hygropause and allow water, usually trapped near the surface, into the middle and upper atmosphere, where it can be photolyzed and ultimately lost from Mars \citep{2024NatAs...8..827K}. \added{Terrestrial dust devils, while impacting regional air quality, have a more modest impact on the Earth's atmosphere \citep{2016SSRv..203...89F}.}

However, we lack basic knowledge about dust devils -- exactly how often dust devils occur and especially how much dust they can lift. The relationships tying dust devil occurrence and physical properties to the ambient conditions in which they form are also unclear, in spite of decades of theoretical, laboratory, and field work \citep{2016SSRv..203..183R}. 

Among the most important questions in dust lifting is what sets the diameter $D$ of a given dust devil. All other things equal, the wider a dust devil, the more dust it lifts into the atmosphere. Presumably, for a given dust flux \added{density (dust lifted per area per time)}, the total dust lifted is the product of the dust devil lifetime (which seems to scale with diameter -- \citealp{2013Icar..226..964L}), the flux (which may scale with the pressure deficit at the devil's center -- \citealp{2010Icar..206..306N}), and the area of the dust devil along the surface (which scales with $D^2$). Observations of dust devils on Mars and Earth show they come in a wide range of sizes and dust-lifting capacities. Apparently, because the martian planetary boundary layer can be many times deeper than Earth's, martian dust devils tend to be much taller and wider: where a typical terrestrial dust devil might span several to ten meters in diameter and hundreds of meters in height \citep{1970JGR....75..531R}, a martian dust devil could be hundreds of meters across \citep{2021PSJ.....2..206J} and kilometers tall \citep{2015Icar..260..246F}. 

For both worlds, though, the larger dust devils occur less frequently. Indeed, the \added{differential} distribution of dust devil diameters seems to fall precipitously for larger and larger values, and several studies have explored this distribution \citep{2009Icar..203..683L, 2010Icar..209..851P, 2011Icar..215..381L}. Since observed diameters span a very wide range, from meters to hundreds of meters or more, the best way to bin and represent the observed distribution is not obvious. \citet{2011Icar..215..381L}, for instance, explored several schemes for creating histograms of different diameter datasets and various functional forms to describe the resulting histograms. As discussed in \citet{2011Icar..215..381L}, the best-fit functional form depends on how the data are binned. Worse still, it is not clear why a particular functional form might arise in nature, though \citet{2011Icar..215..381L} and \citet{2012Icar..219..556K} both highlighted possible physical motivations. \citet{2016SSRv..203..277L} also provided a comprehensive review of dust devil statistics across various dimensions, including dust devil diameter and considerations for creating histograms, among other aspects.

Publication of a large survey of martian dust devils \citep{2025P&SS..25906072C} motivates a fresh assessment of this venerable problem. Folding in recent field work conducted by our team \citep{2025LPSC}, we explore the questions of what determines the diameter of a dust devil and what sets their distribution. In this study, we propose that the diameter distribution arises from close-packing of the dust devils, but not of their visual diameters, rather their areas of influence. We combine this assumption of close-packing with a simple model for dust devil vorticity to work out both the dependence of diameter on ambient conditions and the diameter distribution. Using several different histogramming schemes, we show that observational surveys of dust devil diameters agree with our proposed diameter distribution. We then show that these results predict a specific relationship between dust devil areal density and the so-called dust devil activity index DDA, first proposed in \citet{1998JAtS...55.3244R}. Finally, we discuss important implications of our work and possible future avenues for research.

As explored below, we frame our results in terms of first-order considerations of the vortex kinetics and thermodynamics, but of course, real vortices exhibit richer dynamics than included by our simple, heuristic model. Indeed, \citet{2018Icar..300...97K} and \citet{2019Icar..317..209K} explore some of the complexities we have neglected here. However, given the limited quality of available lab and field data, our model captures the essential features and, as we show below, makes predictions that seem to be supported by the data. As additional data mount, future work should re-visit and expand on the model described here.

\section{Model}\label{sec:Model}

For our model of dust devil dynamics, we follow \citet{1998JAtS...55.3244R}. Based on field observations and theoretical expectations, that study proposed that dust devils can be treated as a thermodynamic heat engine in which sensible heat \added{flux} is absorbed near the planetary surface. This energy then drives convective motions up to the top of the dust devil, where excess thermal energy is radiated into the cool upper atmosphere. Then the flow sinks slowly back down, closing the cycle. 

As part of that model, \citet{1998JAtS...55.3244R} assumed inflow parallel to the surface until the flow reaches the dust devil eyewall, at which point centrifugal acceleration very nearly balances the inward pull of the pressure depression at the dust devil's center. \citet{2020Icar..33813523J} extended this model to provide a scaling relationship for the eyewall angular momentum. These results are the starting point for the present study.

\citet{2020Icar..33813523J} argued that the specific angular momentum (angular momentum per unit mass) $l$ within the eyewall of a dust devil can be derived from the shear in lateral wind speed in the ambient wind field. Thus,
\begin{equation}
    l \sim \upsilon_\theta \left( D/2 \right) \sim \alpha \left( D_{\infty}/2 \right)^2,\label{eqn:angular_momentum}
\end{equation}
where $\upsilon_\theta$ is the azimuthal velocity of the air at the eyewall, $D$ is the eyewall diameter, $\alpha$ is the lateral shear in the ambient wind field, and $D_{\infty}$ is the distance out to which the dust devil draws in air. We will call this parameter the diameter of influence. \added{Equation \ref{eqn:angular_momentum}} allows us to solve for $D_{\infty}$ in terms of the other parameters:
\begin{equation}
    D_{\infty} \sim \left( \frac{2 \upsilon_\theta D }{\alpha} \right)^{1/2}.\label{eqn:D_infty}
\end{equation}

\begin{figure}
    \centering
    \includegraphics[width=\linewidth]{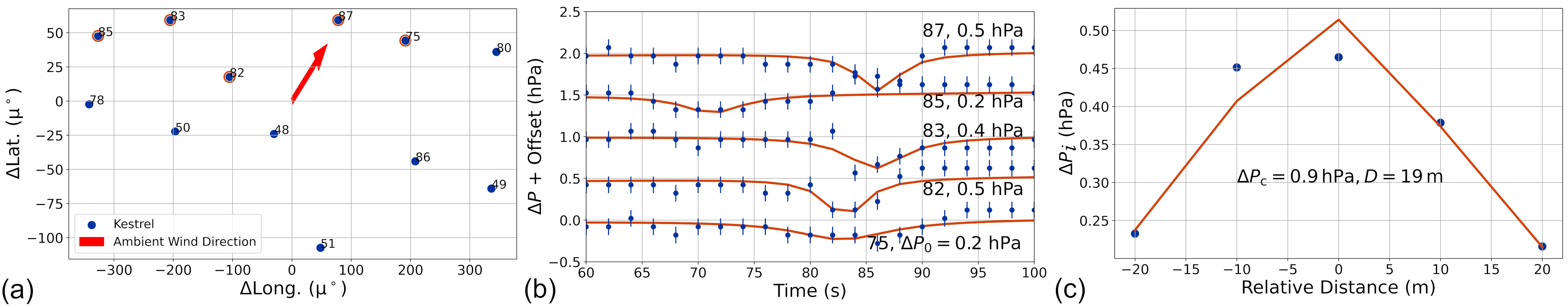}
    \caption{(a) Approximate locations of weather stations in the field. The stations were deployed in a regular grid, but the positions shown exhibit the GPS uncertainties of a few meters. The x axis shows the location along lines of longitude and the y axis along lines of latitude, both in microdegrees ($\mu ^\circ$). Stations that registered the dust devil encounter considered here are shown in orange circles, and the ambient wind vector before the encounter is shown as a red arrow. (b) The pressure time-series for each of the registering sensors shown as blue dots along with the best-fit models, with vertical offsets included for clarity. (The clocks on all the weather stations are not perfectly synchronized, but they do not need to be for this experiment.) The corresponding station numbers (75, 82, 83, 85, and 87) are shown, as well as the maximum pressure excursion for each sensor $\Delta P_i$. (c) The maximum pressure excursion seen by each sensor $\Delta P_i$ as a function of their positions relative to the station with the smallest $b_i$. By fitting Equation \ref{eqn:pressure_excursion}, we estimate the pressure excursion at the center of the dust devil $\Delta P_{\rm c} = 0.9\,{\rm hPa}$ and the diameter $D = 19\,{\rm m}$.}
    \label{fig:Dust-Devil-Encounter_2024Jun24_14-45p}
\end{figure}

How much larger is $D_{\infty}$ than $D$? In other words, how far beyond the visible eyewall do we expect a dust devil's influence extends? Recent field work \citep{2025LPSC} provides a basis for a preliminary assessment. That field work was intended to probe the structures of active dust devils. To that end, a network of a dozen handheld weather stations were deployed in two lines roughly perpendicular to the ambient wind on the Alvord Desert, a playa known for dust devil activity \citep{2018RemS...10...65J}. The stations were separated by about 10 meters from one another (Figure \ref{fig:Dust-Devil-Encounter_2024Jun24_14-45p}(a)). The weather stations, Kestrel 5500s, recorded, among other variables, ambient pressure at 0.5 Hz. On 2024 Jun 24 at about 2:45p local time, a dust devil crossed through the network, registering clear detections in the pressure time-series of several of the weather stations\footnote{A video of the encounter is available here - \url{https://youtu.be/2VkqBUONbbU}.}\added{, similar to the detections reported in \citet{2012GI......1..209L}}. When the dust devil was at its closest point to station $i$, it lay a distance $b_i$ from that station, and the station registered a pressure excursion
\begin{equation}
    \Delta P_i = -\frac{\Delta P_{\rm c}}{1 + \left( \frac{2 b_i}{D} \right)^2}, \label{eqn:pressure_excursion}
\end{equation}
where $\Delta P_{\rm c}$ is the pressure excursion at the center of the dust devil's convective cell. \added{(Equation \ref{eqn:pressure_excursion} is a Lorentzian pressure profile -- \citealp[cf.][]{2010JGRE..115.0E16E}.)} Figure \ref{fig:Dust-Devil-Encounter_2024Jun24_14-45p}(b) shows the $\Delta P_i$ estimated for each station. By folding in the relative placement of all the weather stations, we can fit the above equation to estimate $\Delta P_{\rm c}$ and $D$. Figure \ref{fig:Dust-Devil-Encounter_2024Jun24_14-45p}(c) shows the best-fit values, $\Delta P_{\rm c} = 0.9\,{\rm hPa}$ and $D = 19\,{\rm m}$. Assuming cyclostrophic balance between the centrifugal and pressure accelerations at the eyewall \citep{1998JAtS...55.3244R}, we can convert $\Delta P_{\rm c}$ (and air density $\rho$) into an approximate $v_\theta = \sqrt{\Delta P_{\rm c}/\rho} \approx 9.5\,{\rm m\ s^{-1}}$.

The weather stations also recorded a wind vector (speed and cardinal direction) time-series, and we can use those data to estimate the ambient lateral wind shear before the encounter. (Although the wind speed data can also give us the eyewall wind speed, the winds were quite turbulent during the encounter, and so for this preliminary assessment, we used the cyclostrophic balance assumption to estimate $v_\theta$, instead of a direct estimate. A more comprehensive analysis is left for future work.) To estimate $\alpha$, we calculated the wind vectors from two stations within our network separated by about 40 m along a direction orthongonal to the wind direction. We then averaged those wind vectors over the 10 minutes leading up to the encounter and calculated the magnitude of the vector difference divided by the station separation, giving $\alpha \approx 0.01\,{\rm s^{-1}}$. A similar result derives from consideration of the near-surface winds within a boundary layer convective cell. On an afternoon with active dust devils, wind speeds typically measure $U \sim 5\,{\rm m\ s^{-1}}$ across a convective cell with width $L \sim 1000\ {\rm m}$ comparable in scale to the boundary layer depth \citep{2016SSRv..203..183R}. Thus, we expect $\alpha \sim U/L \sim 0.005\, {\rm s}$, very similar to the estimate from the field data.

Plugging all these estimates into Equation \ref{eqn:D_infty}, we find
\begin{equation}
    D_\infty = \left( \frac{2 \left( 9.5\,{\rm m\ s^{-1}} \right) \left( 19\,{\rm m}\right)}{\left( 0.01\,{\rm s^{-1}} \right)} \right)^{1/2} = 190\,{\rm m}.
\end{equation}
In other words, the inflow for this dust devil appears to extend out to a distance ten times its radius. Now, of course, the exact area of influence for a particular dust devil depends on the specific dust devil parameters and the ambient conditions, but if this instance is characteristic of dust devils generally, we could reasonably conclude that the area of influence within the near-surface atmosphere is much larger (in this case, 100 times larger) than might be apparent from the visual diameter. Some experimental support for this estimate can be drawn from time series meteorological measurements of dust devil encounters \citep{2016Icar..271..326L} where the wind speed and direction are seen to be detectably perturbed from their background values for a duration 5-10 times the full-width half-minimum of the pressure time series (which corresponds to the wall diameter). 

\begin{figure}
    \centering
    \includegraphics[width=0.5\linewidth]{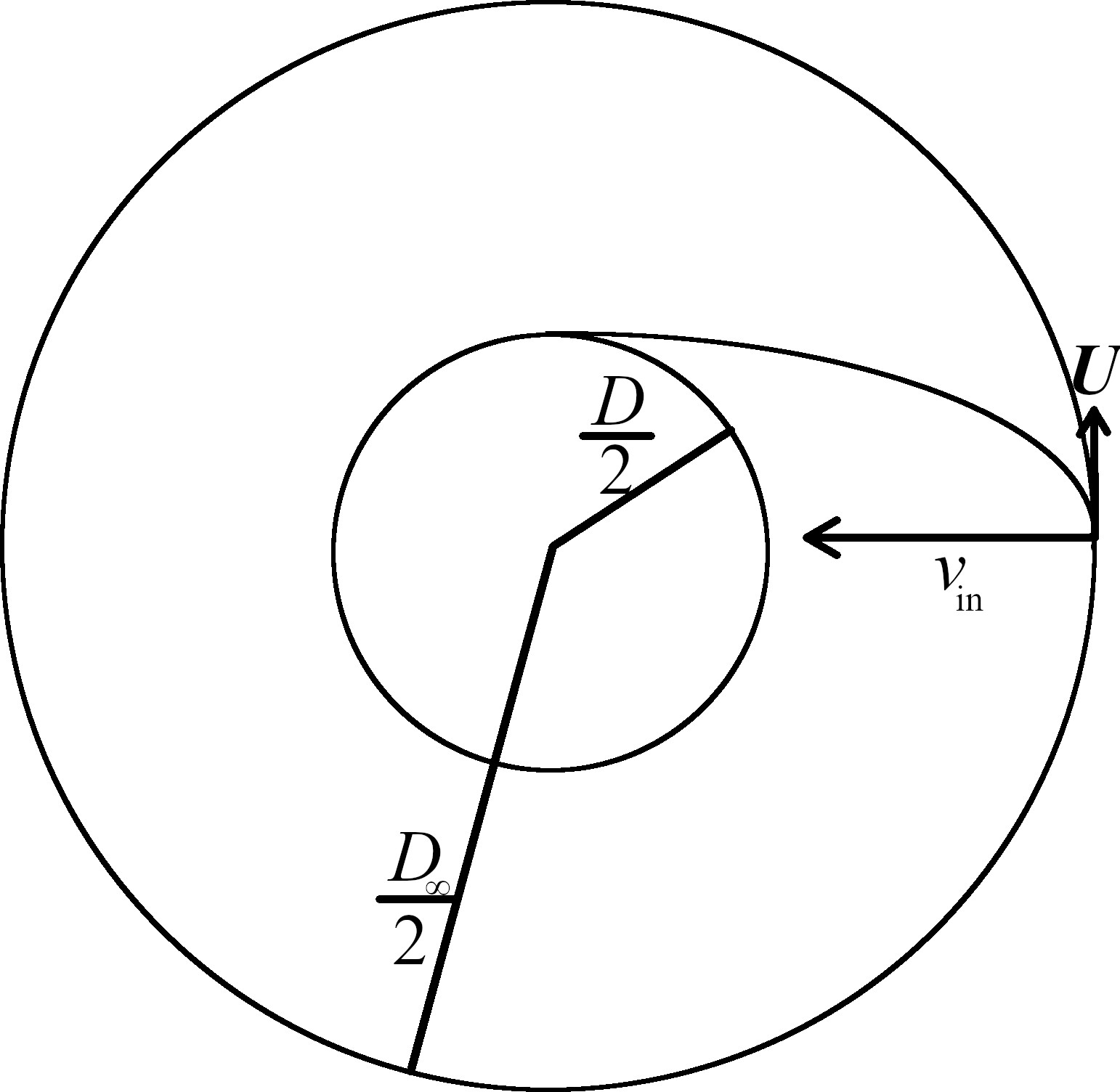}
    \caption{Schematic of dust devil inflow. The innermost circle with radius $D/2$ represents the convective cell at the center of the dust devil. The outermost circle with radius $D_\infty/2$ represents the diameter of influence. The component of the wind perpendicular to the dust devil is given by $U \approx \alpha \left( D_\infty/2 \right)$ where $\alpha$ is the lateral wind shear, and the radial component of the wind is given by $\upsilon_{\rm in}$. An example streamline is also shown.}
    \label{fig:Dust_Devil_Geometry}
\end{figure}

We can also leverage considerations of the inflow geometry to develop an independent relationship between $D$ and $D_\infty$. Figure \ref{fig:Dust_Devil_Geometry} shows the geometry. At a radial distance from the dust devil center $D_\infty/2$ and in a direction perpendicular to the background wind field far from the dust devil, the wind vector has a inward component $\upsilon_{\rm in}$ and a perpendicular component $U$ (as measured relative to the motion of the dust devil center). In order for a streamline to intersect the dust devil, we require that the radially inward distance traveled by an air parcel in time $\Delta t$ is at least $D_\infty/2$ during time $\Delta t = D/U$. Otherwise, the streamline will miss the dust devil and not be incorporated into the inflow. This requirement can be cast as
\begin{equation}
    \left( \frac{U}{\upsilon_{\rm in}} \right) \sim \left( \frac{D}{D_\infty} \right).\label{eqn:inflow_geometry_initial_equation}
\end{equation}
In other words, if $\upsilon_{\rm in}$ were very small and $U$ were very large, the incoming streamline would pass above (as depicted in Figure \ref{fig:Dust_Devil_Geometry}) the dust devil and not join up with the intake. On the other hand, field work \citep{2016SSRv..203..183R} and simulations \citep{2005QJRMS.131.1271K} suggest that streamlines can also orbit a dust devil a handful of times, meaning Equation \ref{eqn:inflow_geometry_initial_equation} is probably not strictly obeyed in real dust devils. However, even for orbiting parcels, it holds approximately. We also expect that the inward component will increase in magnitude as the flow approaches the dust devil, as depicted by the in-spiraling streamline in Figure \ref{fig:Dust_Devil_Geometry}, so, in this sense, Equation \ref{eqn:inflow_geometry_initial_equation} gives an approximate upper limit to the relative velocity components. With our assumption that the background wind field obeys a simple shear relationship, we have $U \approx \alpha \left( D_\infty/2 \right)$, and
\begin{equation}
    \left( \frac{\alpha \left( D_\infty/2 \right)}{\upsilon_{\rm in}} \right) \sim \left( \frac{D}{D_\infty} \right).\label{eqn:inflow_geometry}
\end{equation}

We can use mass conservation to work out another relation. For the convective updraft within the dust devil itself, we have
\begin{equation}
    \dot{M} \sim \pi \rho \left( D/2 \right)^2 w\label{eqn:convective_mass_flux},
\end{equation}
where $\dot{M}$ is the vertical mass flux of air upwards inside the dust devil's eyewall, $\rho$ is the air density (assumed constant), and $w$ is the convective velocity. Since this mass flux comes from the inflow, we can also estimate the horizontal mass flux towards the eyewall passing through a cylinder of radius $D_\infty/2$:
\begin{equation}
    \dot{M} \sim 2 \pi \left( D_\infty/2 \right) \rho H \upsilon_{\rm in} \label{eqn:inflow_mass_flux},
\end{equation}
where $H$ is the height of the topmost inflowing streamline and $\upsilon_{\rm in}$ is the velocity of that streamline. What is the height of the inflow? Observations in the field suggest the inflow is confined to very near the surface, extending to perhaps no more than a few meters in height, where frictional forces dominate. It seem\added{s} plausible that the inflow height would scale with the absolute value of the \added{Monin-}Obukhov length since, at that height and above, convection dominates over shear \citep{2001imm..book.....A}. Field work suggests dust devils are most active when the absolute value of the \added{Monin-}Obukhov length is only a few to ten meters \citep{2016SSRv..203..183R, 2022AeoRe..5900831F}.

Equating the two mass fluxes, incorporating Equation \ref{eqn:inflow_geometry}, and re-arranging gives
\begin{equation}
    \left( \frac{D}{D_\infty} \right) \sim \left( \frac{2 \alpha H}{w} \right)^{1/3}.\label{eqn:inflow_geometry_diameter_ratio}
\end{equation}

We can check this relation by plugging in numbers from the same field work as described above. 
\begin{equation}
    \left(\frac{D}{D_\infty}\right) \sim \left( \frac{2 \left( 0.01\,{\rm s^{-1}} \right) \left( 1\,{\rm m} \right)}{ \left( 10\,{\rm m\ s^{-1}} \right)} \right)^{1/3} \approx 0.13,
\end{equation}
where we have taken a convective velocity $w \approx 10\, {\rm m\ s^{-1}}$ from the field work reported in \citet{1966PhDT........31S} and an inflow height $H \sim 1\,{\rm m}$. Although the values involve considerable uncertainties, the $1/3$ exponent means the results are not all that sensitive to the precise values, and the fact that our theoretical model provides a scaling relationship that closely matches our field-determined value for $D/D_\infty$ provides some reassurance that the model captures the essential features of the system.

Altogether, these results point to a possible explanation for the distribution of dust devil diameters observed in the wild -- convective vortices, of which dust devils are an example, are closely packed. Under this hypothesis, the areas of influence for active vortices cannot overlap because the intake of air for one vortex would disrupt the intake for another. In that case, we might expect the areal density of dust devils (number per unit area) to depend on $D_{\infty}$, and not on $D$, as $D_{\infty}^{-2}$ \citep{2015Icar..260..246F}. This argument may apply more robustly to the formation of dust devils than to their general distribution: once formed, it may be that the velocity field induced by one vortex draws another towards it  (leading to the not-uncommon observation of `companion' or `little brother' dust devils - see, e.g., Figure 4 of \citealt{2016SSRv..203..209K}). 

This dependence on $D_{\infty}$ seems to closely resemble the observed distribution of visual dust devil diameters reported in the literature. For example, \citet{2011Icar..215..381L}  considers several ways to construct the histograms of observed diameters and different analytic representations for those histograms before settling on a power-law distribution with a power-law index of -2, suspiciously close to the closest-packing hypothesis proposed here. The key difference is that that power-law applies to the distribution of visual diameters $D$ and not the diameter of influence considered here $D_{\infty}$. As we discuss below, though, a power-law closer to $D^{-5/3}$ is both what we expect from the hypothesis here and what the data seem to support.

Given the assumption that dust devils are closely packed in $D_\infty$, what distribution in $D$ do we expect? Assuming the histogram of dust devils in $D_\infty$ is given by $dN/dD_\infty \propto D_{\infty}^{-2}$, we can work out the histogram in $D$ as
\begin{equation}
    \frac{dN}{dD} = \left( \frac{dN}{dD_\infty} \right) \left( \frac{dD_\infty}{dD} \right).\label{eqn:dN_dD}
\end{equation}

We can use Equation \ref{eqn:D_infty} to calculate $dD_\infty/dD$. However, there is another dependence on $D_\infty$ implicit in $v_\theta$ that must be included. \citet{1998JAtS...55.3244R} argued that the work done by the pressure acceleration into the dust devil was balanced by the work lost to friction along the inflow streamlines:
\begin{equation}
    \int_{r = D_\infty/2}^{0} \frac{dP}{\rho} \approx \int_{r = D_\infty/2}^{0} \vec{f}\cdot d\vec{l},\label{eqn:work_integral}
\end{equation}
where $\vec{f}$ is the friction force per unit mass along some point in the intake streamline, $d\vec{l}$ is an infinitesimal distance along that streamline, and the integral is taken a radial distance $D_\infty/2$ to very near the dust devil's center. 

According to the ideal gas law, $dP/\rho = R T d \ln P$, where $R$ is the gas constant and $T$ is the fluid temperature. The left side of the integral evaluates to $R T_{\rm s} \ln\left( \left( P_{\rm s} - \Delta P_{\rm c}\right)/P_{\rm s}\right)$, where $T_{\rm s}$ is the (assumed constant) near-surface air temperature and $P_{\rm s}$ is the surface pressure. $\Delta P_{\rm c} \ll P_{\rm s}$, so 
\begin{equation}
    \ln\left( \frac{P_{\rm s} - \Delta P_{\rm c}}{P_{\rm s}} \right) \approx -\frac{\Delta P_{\rm c}}{P_{\rm s}},\label{eqn:approx_Delta_P}
\end{equation}
and $P_{\rm s}/R T_{\rm s} = \rho$. Folding this result into Equation \ref{eqn:work_integral}, we see
\begin{equation}
     \frac{\Delta P_{\rm c}}{\rho} \approx f D_\infty/2,
\end{equation}
where $f$ is the frictional acceleration averaged over the inflow streamline. \added{($\vec{f} \cdot d\vec{l} < 0$, which cancels the minus sign on the right-hand side of Equation \ref{eqn:approx_Delta_P}.)} Invoking cyclostrophic balance gives 
\begin{equation}
v_\theta = \sqrt{\Delta P_{\rm c}/\rho} \approx \sqrt{f D_\infty/2}.
\end{equation}

Plugging this expression into Equation \ref{eqn:D_infty} and solving for $D_\infty$ gives
\begin{equation}
    D_\infty \sim \left( \frac{2 f D^2}{\alpha^2}\right)^{1/3}.\label{eqn:relating_Dinf_to_D}
\end{equation}
Then implementing this expression into Equation \ref{eqn:dN_dD} gives the following scaling for the histogram of dust devil diameters $D$:
\begin{equation}
    \frac{dN}{dD} \propto D_\infty^{-2} \left( \frac{dD_\infty}{dD} \right) \propto D^{-5/3}.
\end{equation}
As we show in the next section, data from several surveys\added{, particularly for the smallest, most numerous dust devils,} match this prediction.

\added{It is worth highlighting here that some of the environmental variables, such as $\alpha$ and $\vec{f}$, are likely to vary from between dust devils, locations, times of day, etc. and that they may also exhibit dependence on (or at least correlation with) the dust devil parameters themselves. For example, if the frictional force scaled with intake velocity, which seems plausible, the $dN/dD$ scaling proposed here would have to be modified and the index $-5/3$ might take on a different value. As reiterated in Section \ref{sec:Conclusions}, the scaling relations here are intended to be approximate and applicable in an average sense, not necessarily quantitatively predictive, devil to devil.}

\section{Observed Dust Devil Diameters} \label{sec:Observed_Dust_Devil_Diameters}

As discussed in the Introduction, determining the best and most meaningful representation for the distribution of dust devil diameters has been the subject of considerable exploration. Far from being simply an intellectual exercise, the distribution of diameters plays an important role in estimates for the contribution of dust devils to a planetary dust budget. A distribution that skews more toward larger devils implies more dust lifted in a population-weighted sense. In an extensive review, \citet{2016SSRv..203..277L} discussed many of the issues related to dust devil diameters, among other statistical considerations. That study explored the impact of the chosen histogramming scheme and the number of dust devils included in an analysis on the inferred functional representation. Following on that study, here we consider several dust devil surveys and different histogramming schemes in an attempt to circumvent some of the biases discussed in that and other studies.

\subsection{Comparison to the Diameter Distribution from \citet{2025P&SS..25906072C}}
\begin{figure}
    \centering
    \includegraphics[width=\linewidth]{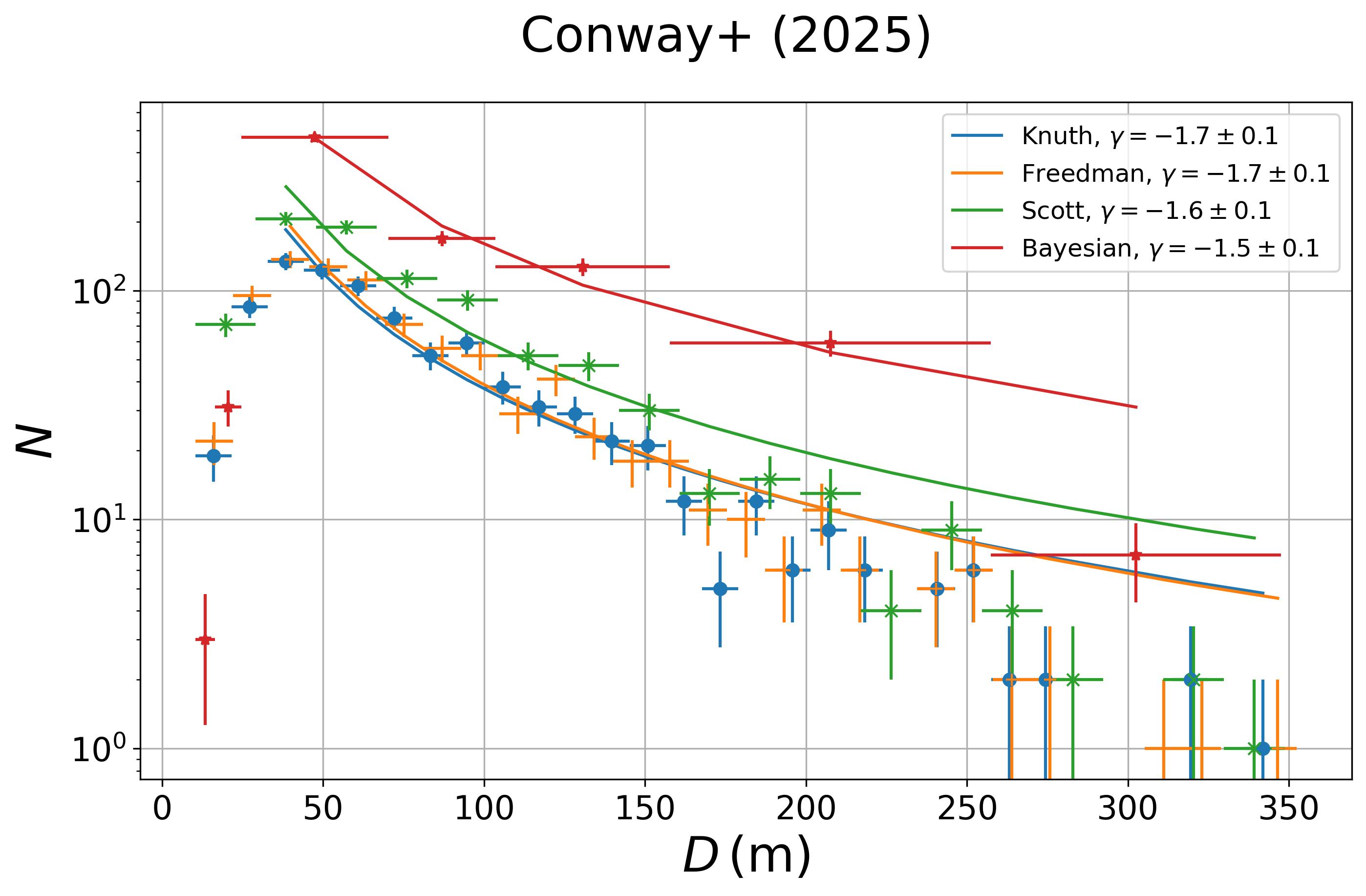}
    \caption{Histogram of dust devil diameters as reported in \citet{2025P&SS..25906072C}. The different color and marker-style points represent histograms created using various binning algorithms as indicated in the legend text and described in the narrative. The horizontal bars on each marker show the width of each bin, and the vertical bars show the Poisson uncertainties. The best-fit power-law for each histogram is shown in the color corresponding to the marker color, and the power-law index $\gamma$ (and uncertainty) for each fit is shown in legend.}
    \label{fig:Conway_Diameter_Histogram}
\end{figure}

\citet{2025P&SS..25906072C} recently published an extensive catalog of dust devils imaged at a resolution of $\sim 6 {\rm m\ pixel^{-1}}$ by the Mars Reconnaissance Orbiter (MRO) Context Camera (CTX). That study used a machine-learning (ML) algorithm to scan 132,359 CTX images from Mars Years 28–36 to yield 13,409 detections of active dust devils. The study included a dataset of manual measurements by citizen scientists through the Zooniverse project. The results of the ML survey were among the most voluminous of any remote-sensing dust devil study and allow comprehensive assessment of regional and seasonal occurrence rates for dust devils across several years. However, the key parameter required for the present study, dust devil diameter, was not directly measured by the ML algorithm. The Zooniverse effort, though, did report size measurements for 862 dust devils. These measurements estimate the widths of dust plumes at the top of the dust devils, as seen from the spacecraft's downward-looking perspective from orbit. Such plumes are likely to be broader and more dispersed than the near-surface dust devil diameters considered here. We assume that the plume widths aloft are linearly proportional to the near-surface diameter $D$ \citep{2025P&SS..25906072C}, and it is these data that we use for the present study to construct  our histograms.

A key question in constructing a histogram of any dataset is how to bin up the data, and a flood of ink has been spilt in devising rules of thumb, heuristics, and algorithms. Each approach has its own advantages and limitations \citep{2012msma.book.....F}. For the present study, in lieu of a comprehensive assessment of the best approach, we consider several readily available schemes and compare the resulting power-law fits. 

Figure \ref{fig:Conway_Diameter_Histogram} shows the histograms. The binning schemes we used are as follows:
\begin{itemize}
    \item Knuth's rule -- Knuth's rule uses bins of fixed-width based on a Bayesian approach to determining the optimal bin width \citep{2006physics...5197K}. This histogram is shown as blue circles, and the bin size for Figure \ref{fig:Conway_Diameter_Histogram} is 11 m.
    \item Freedman's rule -- An optimal and fixed bin width is determined on the basis of dividing the data up into quartiles and the total number of data points \citep{Freedman1981}. This histogram is shown as orange crosses, and the bin size for Figure \ref{fig:Conway_Diameter_Histogram} is 12 m.
    \item Scott's rule -- An optimal and fixed bin width is determined based on the standard deviation of the data and the total number of data points \citep{ScottsRule}. This histogram is shown as green stars, and the bin size for Figure \ref{fig:Conway_Diameter_Histogram} is 19 m.
    \item Bayesian Blocks -- This approach involves iteratively choosing bins to both minimize the variances within bins but also limit the total number of bins on the assumption that, all other things equal, the fewer the bins, the better \citep{2013ApJ...764..167S}. The resulting bins are not necessarily uniform in width, as can be seen in Figure \ref{fig:Conway_Diameter_Histogram}. This histogram is shown as red triangles, and the seven variable bin sizes for Figure \ref{fig:Conway_Diameter_Histogram} range from 6 to 100 m and average 48 m.
    
\end{itemize}

\added{Not surprisingly, the exact number of points within each bin differs from scheme to scheme. For instance, Figure \ref{fig:Conway_Diameter_Histogram} shows that, because of the larger bin sizes, the histogram for the Scott rule (red triangles) has more diameters in many of the bins than the histograms for the other rules. However, the key aspect we are interested in here is the power-law index for the histograms, which depends on the slope of the histogram and not on the exact number of elements in each bin. Therefore, the mismatch in number of binned elements between the rules is not crucial.}

Running from large to small diameters, the number of dust devils in each bin mostly increases up to a diameter of about 50 m, after which point the number of dust devils falls off for all binning schemes. Though we do expect dust devils exhibit a minimum diameter (as discussed later), 50 m is probably larger than that minimum -- certainly, smaller dust devils have been observed from the ground by landed spacecraft \citep[e.g.,][]{2006JGRE..11112S09G}. As described in \citet{2025P&SS..25906072C}, the images analyzed have a ground sampling of $\sim 6 {\rm m\ pix^{-1}}$. We would expect dust devils with diameters less than a few pixels across would be difficult to detect and measure. We assume, therefore, that the decline in devils below the peak in each histogram results from this detection bias, rather than an actual decline in the population, and we only fit power-laws to the larger dust devils.

Another key consideration in our analysis is the inclusion of uncertainties in each bin. We assume Poisson uncertainties -- a bin containing $N$ devils is assigned an uncertainty $\sqrt{N}$. We found that, if we do not include these uncertainties, the best-fit power-law indices are $\gamma \le -2$, inconsistent with our expectations. Not including the uncertainties means each bin is weighted equally in the fit, regardless of the number of devils in the bin. Under this approach, bins with only a single dust devil would have as much impact on the final fit as bins with hundreds of devils.

To each of the histograms shown in Figure \ref{fig:Conway_Diameter_Histogram}, we applied linear regression to determine a power-law fits $N(D) \propto D^{\gamma}$ by fitting slopes $\gamma$ and intercepts $b$ using
\begin{equation}
    \log_{10} \left( N(D) \right) = \gamma \log_{10} \left( D \right) + b.\label{eqn:power_law_fit}
\end{equation}
The legend in Figure \ref{fig:Conway_Diameter_Histogram} shows resulting power-law indices for each fit $\gamma$, along with the corresponding uncertainties -- the values all agree to within uncertainties with our expectation of $-5/3 = -1.\bar{6}$ from Section \ref{sec:Model}.

\added{Though the $\gamma = -5/3$ power-law is formally consistent with the best-fit models for this dataset, there may be some mismatch for the diameters of the largest dust devils. As can be seen in Figure \ref{fig:Conway_Diameter_Histogram}, the colored lines representing the model fits seem to over-predict somewhat the number of diameters in the largest bins ($D > 200\,{\rm m}$). Given the small number of points in those bins, this mismatch might simply arise from small-number statistics, but it is also possible that alternative forms for the diameter distribution may apply to the largest dust devils \citep{2006GeoRL..3319S06K}.}

\subsection{Comparison to the Diameter Distributions from Other Studies}
As discussed in Section \ref{sec:Introduction}, \citet{2011Icar..215..381L} conducted a broad survey of the statistics of dust devil diameters based on a variety of studies. That study points to several surveys that report diameters, but, unfortunately, the vast majority of those surveys provide only the final histograms of diameters and not the tally of diameters themselves; moreover, those other surveys report many times fewer devils than in \citet{2025P&SS..25906072C}. Thus, comparison of results from those studies to the results here is difficult. However, we attempt that comparison anyway. Not surprisingly, the reduced number of devils in the other surveys and the pre-histogramming means we derive power-law fits that diverge somewhat from the results from our analysis of the \citet{2025P&SS..25906072C} data.

\begin{figure}
    \centering
    \includegraphics[width=\linewidth]{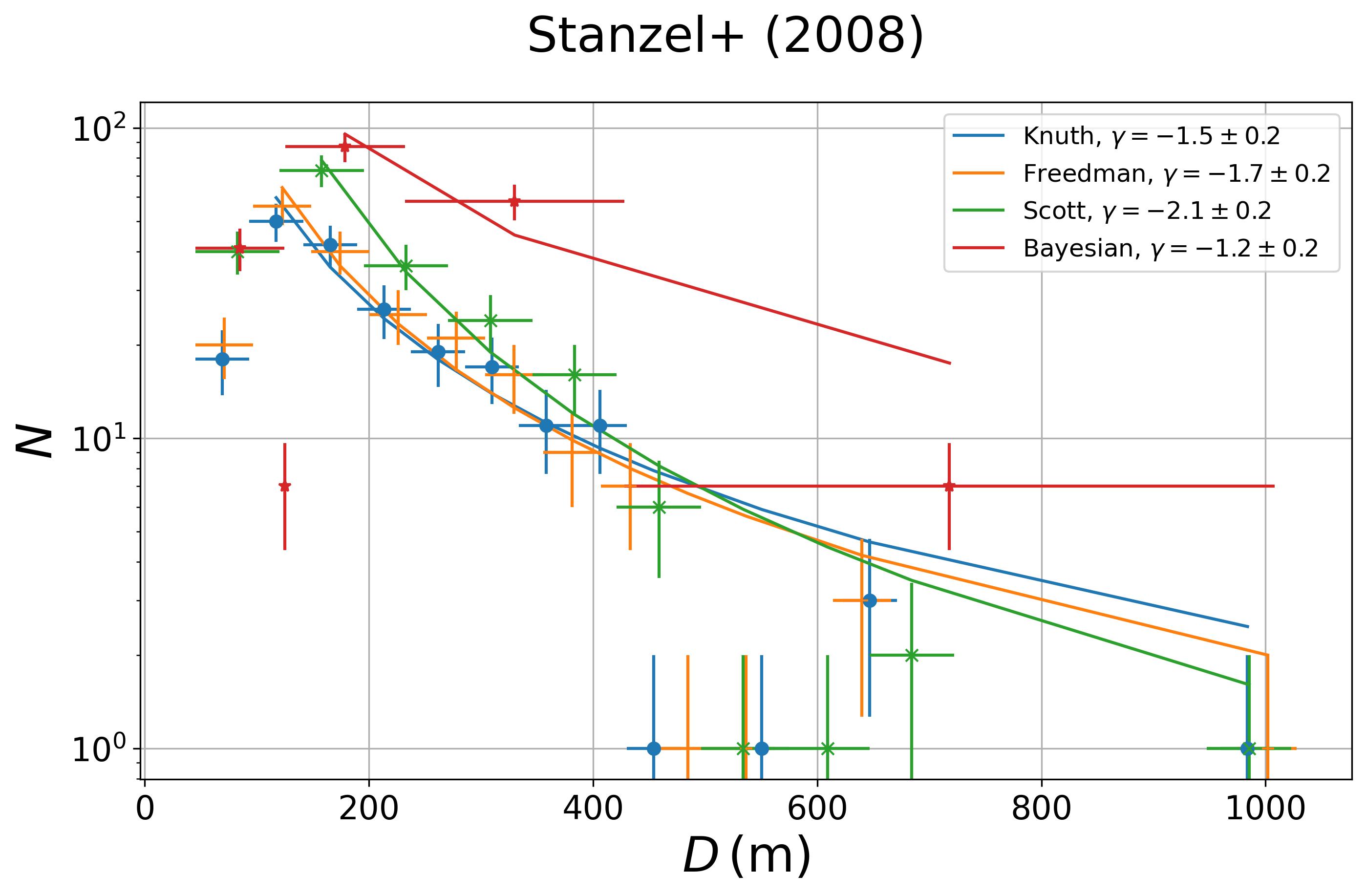}
    \caption{Histogram of dust devil diameters as reported in \citet{2008Icar..197...39S}. The symbols and legend all are defined in the same ways as in Figure \ref{fig:Conway_Diameter_Histogram}.}
    \label{fig:Stanzel_Diameter_Histogram}
\end{figure}

\citet{2008Icar..197...39S} conducted a manual search of 23 High Resolution Stereo Camera (HRSC) images from the Mars Express mission from between January 2004 and July 2006 and yielding a total of 205 dust devils with measured diameters and heights. The images had a resolution of about $25\, {\rm m\ pixel^{-1}}$. Figure \ref{fig:Stanzel_Diameter_Histogram} shows the histograms resulting from the same binning schemes as used in Figure \ref{fig:Conway_Diameter_Histogram} (the bin sizes for each scheme are slightly different from the previous bin sizes). As shown in the figure legend, the power-law indices differ somewhat from those in Figure \ref{fig:Conway_Diameter_Histogram}, depending on the exact binning scheme, but all the best-fit $\gamma$ values here agree with the results reported in Figure \ref{fig:Conway_Diameter_Histogram} to within $2\sigma$.

\begin{figure}
    \centering
    \includegraphics[width=\linewidth]{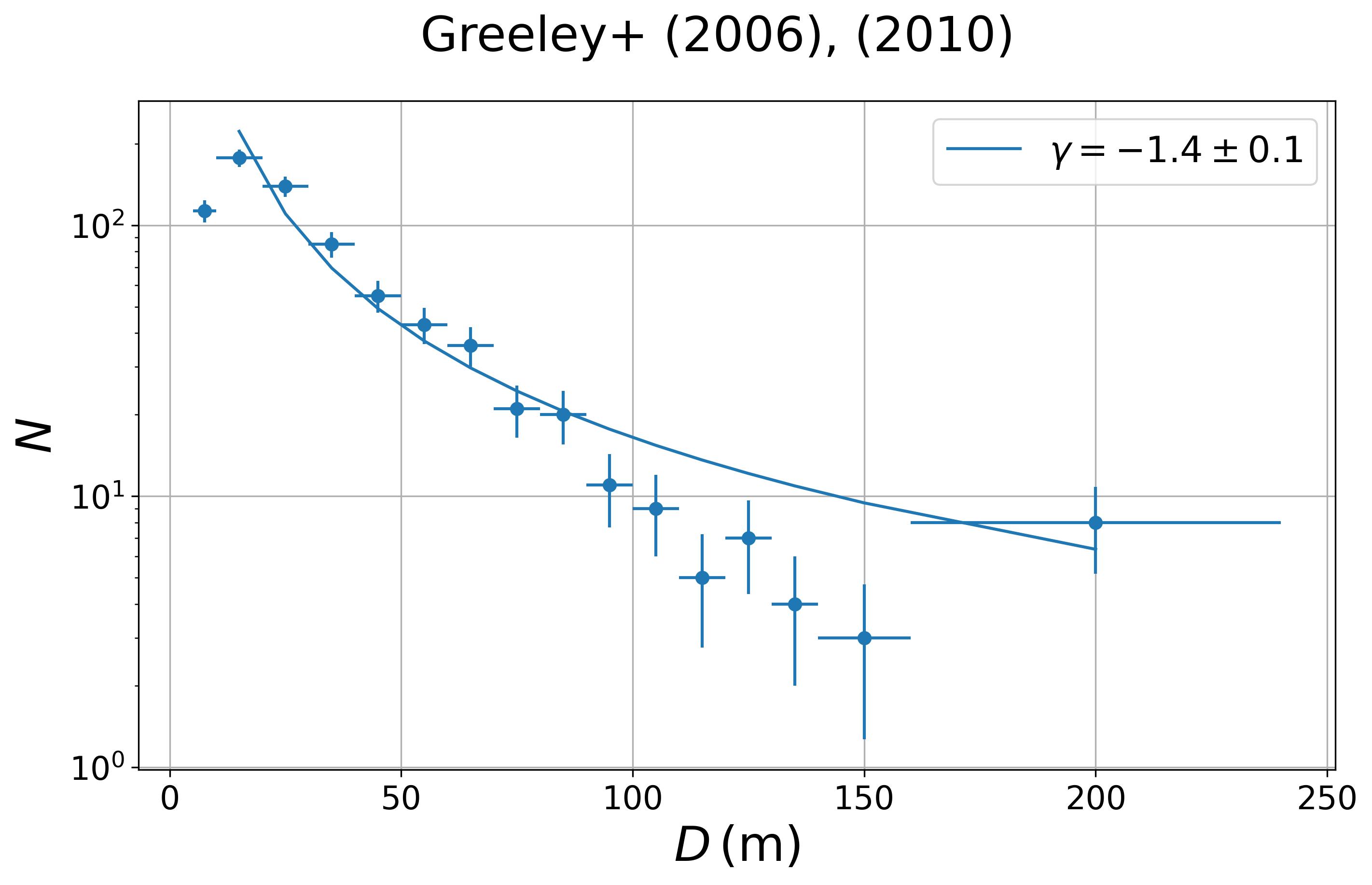}
    \caption{The histogram of dust devil diameters reported in \citet{2006JGRE..11112S09G} and \citet{2010JGRE..115.0F02G}. Unlike Figures \ref{fig:Conway_Diameter_Histogram} and \ref{fig:Stanzel_Diameter_Histogram}, we do not have the diameters directly, only a histogram of values. Thus, the binning shown here does not follow any of the schemes used in the previous figures.}
    \label{fig:Greeley_Diameter_Histogram}
\end{figure}

\citet{2006JGRE..11112S09G} and \citet{2010JGRE..115.0F02G} conducted a dust devil survey using images collected by the Spirit rover between March 2005 and October 2009, netting 736 dust devils with measured diameters. Since the imaged dust devils ranged in distance from the rover, the images did not have a fixed linear resolution, but not surprisingly, the number of dust devils with smaller diameters appears to decline, likely because of the limited image resolution. Unfortunately, that study did not report the diameters directly but only a histogram of diameters; thus, we cannot apply the binning schemes used in the previous figures. Figure \ref{fig:Greeley_Diameter_Histogram} shows the histogram and our power-law fit. Again, not surprisingly, the power-law index differs somewhat from the other indices but does agree to within $2\sigma$. \added{As pointed out above for Figure \ref{fig:Conway_Diameter_Histogram}, the power-law fit in Figure \ref{fig:Greeley_Diameter_Histogram} does seem to over-predict somewhat the numbers of larger dust devils, which, again, may be the result of a different distribution applying for those largest diameters.}

\section{Discussion}\label{sec:Discussion}
\added{In this section, we expand on the relationships developed in previous sections. In particular, we show that the distribution of dust devil diameters proposed above implies that the areal density of dust devils (number per unit area) scales directly with the product of their thermodynamic efficiency and the sensible heat flux at the surface (Equation \ref{eqn:number_density_scales_with_DDA} below). This product is called the ``dust devil activity'' DDA in \citet{1998JAtS...55.3244R}. First, t}hough the relationships developed here between the various dust devil parameters are novel, many of the concepts have been explored in previous work. Already mentioned, \citet{2011Icar..215..381L} considered the statistics of dust devil diameters, specifically with an eye toward how best to bin up the diameter tallies and what functional form to fit to the resulting histograms. Among other points made, \citet{2011Icar..215..381L} argued that a power-law seemed to be the most effective and physically motivated fit for the histograms. A power-law index $\gamma = -2$ was considered but not specifically fit to the histograms of dust devil diameters $D$. Of course, our analysis here suggests a power-law $\propto D_\infty^{-2}$ and $\propto D^{-5/3}$.

\citet{2015Icar..260..246F} conducted a survey of dust devils using 127 CTX images collected between November 2006 and July 2013, netting more than 2000 active dust devils. However, that study only measured the heights of the dust devils and the relative spacings and not their diameters. The focus in that study was to correlate the spacings with the heights to explore the hypothesis that both scaled with the planetary boundary layer (PBL) thickness. Among other results, the study found that the martian PBL thickness is $\sim 5$ times the observed median dust devil heights and suggested that the average distance between nearest-neighbor dust devils is $\sim1-2$ times the PBL thickness. 

The results in the present study do not bear on the PBL thickness since we did not analyze any heights, but they could be related to the dust devil spacing. The key hypothesis in the present study is that dust devils are closely packed in $D_\infty$ but not necessarily in $D$. \citet{1979JGR....84..295W} suggested that dust devils form preferentially at the vertices of hexagonal (or at least, polygonal) regional convective cells, whose sizes scale with PBL thickness. However, the hypothesis here that dust devils are maximally packed in $D_\infty$ does not obviously contradict that suggestion -- dust devils will occur where conditions are conducive, and so the maximal packing hypothesis applies only to those regions and not everywhere in the boundary layer. Thus, we might expect that, if dust devils really are confined to the vertices of larger convective cells, they could both be spaced in a way that scales with PBL thickness and a way that maximally packs regions with conducive conditions.

Our results here also allow us to predict the relationship between energy available to power dust devils and the areal density of dust devils by considering the fractional area occupied by convective motions within dust devils. As discussed above, \citet{1998JAtS...55.3244R} suggested dust devil inflows run parallel to the surface until they enter the eyewall, at which point they ascend. In the parlance of our model, the vertical updraft associated with the dust devil takes place only within a horizontal area $\pi \left( D/2 \right)^2$, while the entire system occupies a surface area $\pi \left( D_\infty/2 \right)^2$. Further, within the model outlined in \citet{1998JAtS...55.3244R}, the thermodynamic energy that drives dust devil activity is given by ${\rm DDA} = \eta F_{\rm s}$, where $\eta = \left( T_{\rm hot} - T_{\rm cold}\right)/T_{\rm hot}$, $T_{\rm hot}$ is the temperature of the air at the center of the dust devil, and $T_{\rm cold}$ is the temperature of the atmosphere at the top of the dust devil's ascent. Thus, $\eta$ represents the thermodynamic efficiency of the dust devil heat engine. $F_{\rm s}$ is the sensible heat flux that warms and makes buoyant the inflowing air. 

Under the same framework, \citet{1996JAtS...53..572R} showed that the fractional area occupied by convective motions within a convective system is given by 
\begin{equation}
    \sigma \approx \mu^{1/2} \left( \frac{8 \epsilon \sigma_{\rm R} T_{\rm cold}^3}{\rho c_p} \right)^{3/2} \left( \frac{\eta F_{\rm s}}{\rho} \right)^{-1/2},\label{eqn:fractional_convective_area}
\end{equation}
where $\mu$ is a dimensionless coefficient of turbulent dissipation of mechanical energy, $\epsilon$ is the atmospheric emissivity, $\sigma_{\rm R}$ is the Stefan-Boltzmann constant, and $c_p$ is the specific heat capacity at constant pressure. Thus, we expect the fractional area occupied by convective motions to scale as $\left( \eta F_{\rm s} \right)^{-1/2}$. It is worth noting that dust devils are not the only form of convection occupying surface area, of course, and they are not even the only mechanism by which dust is raised \citep{NEWMAN2022637}. We return to this point below, but what does this expectation about fractional area mean about how $\eta F_{\rm s}$ is related to dust devil areal density?

Assume that the areal density occupied by dust devils per bin in $D_\infty$ is given by
\begin{equation}
    \frac{dN}{dD_\infty} = N_0 f_{D_\infty} D_\infty^{-2},
\end{equation}
where $N_0$ is the total areal density of dust devils integrated over all bins and $f_{D_\infty}$ is the normalization constant defined such that
\begin{equation}
    1 = f_{D_\infty} \int_{D_\infty^{\rm min}}^{D_\infty^{\rm max}} D_\infty^{-2} dD_\infty \Rightarrow f_{D_\infty} \approx D_\infty^{\rm min},
\end{equation}
and we have assumed the maximum value $D_\infty^{\rm max} \gg D_\infty^{\rm min}$. This assumption is justified on the basis of Equation \ref{eqn:D_infty}: $D_\infty \propto D^{2/3}$, assuming $f$ and $\alpha$ are independent of $D$ \added{(though they may not be -- see Section \ref{sec:Model})}. Looking at the histograms in Figures \ref{fig:Conway_Diameter_Histogram}, \ref{fig:Stanzel_Diameter_Histogram}, and \ref{fig:Greeley_Diameter_Histogram}, the minimum $D \sim 10\, {\rm m}$, while the maximum $D \sim 1\,{\rm km}$, so $D_\infty^{\rm max}/D_\infty^{\rm min} \sim \left( 1\,{\rm km}/10\,{\rm m}\right)^{2/3} \approx 20$. 

The fractional area occupied by all vertical updraft motions within all dust devils across all $D_\infty$ bins is then given by
\begin{equation}
    \sigma = N_0 D_{\infty}^{\rm min} \int_{D_\infty^{\rm min}}^{D_\infty^{\rm max}} D_\infty^{-2} \pi \left( D/2 \right)^2 dD_\infty
\end{equation}
since vertical updraft within a dust devil system only takes places within diameter $D$ and not throughout $D_\infty$. Using Equation \ref{eqn:relating_Dinf_to_D}, we can replace $D$:
\begin{equation}
    \sigma = \left( \frac{\pi\alpha^2}{8f} \right) N_0 D_{\infty}^{\rm min} \int_{D_\infty^{\rm min}}^{D_\infty^{\rm max}} D_\infty^{-2} D_\infty^{3} dD_\infty \approx \left( \frac{\pi\alpha^2}{16f} \right) N_0 D_{\infty}^{\rm min} \left( D_{\infty}^{\rm max} \right)^2,
\end{equation}
where we have again assumed $D_{\infty}^{\rm max} \gg D_{\infty}^{\rm min}$. How to evaluate the maximum and minimum $D_{\infty}$ values? \added{\citet{1990JApMe..29..498H} suggested that the typical diameter for a population of dust devils is approximately twice the absolute value of the \added{Monin-}Obukhov length $|L|$}, and \citet{2022AeoRe..5900831F} measured $|L|$ values during high dust devil activity of a few to tens of meters. \added{Since the distribution of diameters is significantly skewed toward smaller values, the typical or average diameter also skews toward smaller values. Thus, the typical diameter is comparable to (though, of course, somewhat larger than) the smallest diameter.} On the other end, the largest diameters have been suggested to be some fraction of the planetary boundary layer depth \citep{2021AdSpR..67.2219L, 2016SSRv..203..277L}, hundreds of meters. Of course, these diameters $D$ are not necessarily the same as the diameters of influence $D_\infty$ considered here, but as we show next, we do not require the actual maximum and minimum values to evaluate the scaling relation between dust devil activity and occurrence.

We also expect the fractional convective area relation given in Equation \ref{eqn:fractional_convective_area} to apply dust devil by dust devil since they each are assumed to be approximately closed systems. The fractional convective area for each dust devil is given by 
\begin{equation}
    \sigma = \left( D/D_\infty \right)^2 = \left( \frac{\alpha^2}{2 f} \right) D_\infty \propto \left( \eta F_{\rm s} \right)^{-1/2},
\end{equation}
a scaling we expect should apply for the minimum and maximum $D_\infty$ values. Therefore,
\begin{equation}
    \sigma \propto N_0 D_{\infty}^{\rm min} \left( D_{\infty}^{\rm max} \right)^2 \propto N_0 \left( \eta F_{\rm s} \right)^{-3/2}. 
\end{equation}
In other words, the fractional area occupied by upward motions within a population of dust devils ought to scale inversely with the square root of $\eta F_{\rm s}$ cubed. 

But Equation \ref{eqn:fractional_convective_area} says that, as a whole, the fractional area occupied by convection (dust devils plus all other convective motions) should scale inversely as the square of $\eta F_{\rm s}$ itself. Since the population of dust devils supplies at least some of that fractional coverage, we expect that 
\begin{equation}
    N_0 \left( \eta F_{\rm s} \right)^{-3/2} \propto \left( \eta F_{\rm s} \right)^{-1/2} \Rightarrow N_0 \propto \left( \eta F_{\rm s} \right)\label{eqn:number_density_scales_with_DDA}
\end{equation}
In other words, the closest packing hypothesis suggests the areal density of dust devils should scale directly with the dust devil activity ${\rm DDA}$ given in \citet{1998JAtS...55.3244R}.

This result appears to be borne out by meteorological monitoring and modeling at the martian surface. For example, using data from Mars Science Laboratory, \citet{2019JGRE..124.3442N} identified the passage of vortices and dust devils through analysis of the REMS meteorological data and Navcam/Mastcam imagery. \citet{2019JGRE..124.3442N} tallied up more than three Mars years of such encounters, exploring the dependence of encounter rate on time of day, season, and location, among other possible controlling factors. That study also conducted modeling using the MarsWRF package \citep{2007JGRE..112.9001R, 2012Icar..221..276T, 2015Icar..257...47N, 2019JGRE..124.3442N} and tailored to the specific circumstances of the vortex encounters to assess the meteorological conditions under which the encounters took place. An estimate of DDA was among the meteorological variables calculated in the model, and \citet{2019JGRE..124.3442N} compared these DDA values with the vortex encounter rates, hour-by-hour and season-by-season. The encounter rates seemed to be more sensitive to the modeled DDA estimates than to other possible controls, such as the boundary layer thickness or sensible heat flux alone.

\section{Conclusions}\label{sec:Conclusions}
In this study, we have explored the hypothesis that dust devils occur in a close-packed configuration in regions where conditions are conducive to their formation. The packing is not close as measured by their diameters $D$ but instead by their diameters of influence $D_\infty$; that is, the distances out to which air and angular momentum are drawn in by the convective intake. We have shown that this hypothesis predicts that the distribution of dust devil diameters should follow a $D^{-5/3}$ power-law distribution, and a careful analysis of observed diameters comports with that prediction. 

We have also shown that this close-packing hypothesis directly relates the number of dust devils per area, the dust devil areal occurrence rate $N_0$, to the thermodynamic efficiency of convection $\eta$ and the surface sensible heat flux $F_{\rm s}$ as $N_0 \propto \eta F_{\rm s}$. Although a variety of studies have (very reasonably) assumed a positive correlation between the two, there has been, so far as we know, no work that has worked out the exact scaling. Our analysis appears to be the first to make that direct connection.

Of course, our study makes many important simplifying assumptions that may limit its accuracy. For instance, we have assumed that dust devils derive their vorticities from shear in the ambient wind field. This assumption, used in \citet{2020Icar..33813523J}, likely neglects complexities in the dynamics, and confirmation of the source of dust devil vorticity remains unclear \citep{2016SSRv..203..183R}. We have also assumed the ambient shear and frictional forces within the flow can each be quantified by a single number for all dust devils within an observed population. We expect, in reality, that these aspects of dust devil formation depend on highly localized conditions and time-of-day, and so they probably vary from devil to devil, even for devils originating on the same field site. Even so, we expect that there are reasonable, population-weighted average values for the formation parameters. Additional data and modeling could help develop more precise approaches than used here.

Another key assumption is that dust devils can be characterized with specific, fixed parameters -- $\Delta P_{\rm c}$, $D$, etc. Dust devil vortices have traditionally been modeled using quasi-steady, well-defined pressure, temperature, and wind profiles \citep{2016Icar..271..326L, 2016SSRv..203..209K}, but they are dynamical structures that can grow and evolve as they travel from one area to another. Here, again, though, we argue that the parameters we have considered do not necessarily represent instantaneous values but rather some kind of time-averaged values, marginalized over the lifetime of a given devil. 

The simplicity of our approach enhances its utility. Field and remote-sensing studies of dust devils face numerous technical and logistical challenges that limit their measurements. As already highlighted here, the limited resolution of remote-sensing data introduces important detection biases. Field work sometimes involves imaging of indistinct dust devils at unknown distances \citep{2012Icar..221..632B, 2022AeoRe..5900831F}, which makes determining their physical parameters, including diameters, difficult. Occam's razor dictates that the precision of the model should be commensurate with the precision of the data. 

Even with these limitations, our hypothesis makes additional testable predictions. For example, based on \citet{1996JAtS...53..572R}, we have argued that $\left(D/D_\infty\right)^2 \propto \left( \eta F_{\rm s} \right)^{-1/2}$. Using Equation \ref{eqn:relating_Dinf_to_D}, we can show that $D \propto \left( \eta F_{\rm s} \right)^{-3/4}$. In other words, dust devils should exhibit larger diameters as DDA drops (assuming a sufficiently large DDA to allow their formation). This counterintuitive result arises from the fact that the convective velocity increases as $\left( \eta F_{\rm s} \right)^{1/2}$ \citep{1996JAtS...53..572R}, while the flux of kinetic energy up the dust devil column scales with $w^3$. Thus, the energy available to drive convective motion scales with $\eta F_{\rm s}$, but the energy throughput for a given diameter scales with $\left(\eta F_{\rm s}\right)^{3/2}$. Since DDA is generally a maximum near mid-day and falls off either earlier or later in the day \citep[e.g.,][]{2019JGRE..124.3442N}, we might expect larger dust devils earlier or later in the day. 

Tentative evidence supporting this inference comes from the survey reported in \citet{2021PSJ.....2..206J}. That study used wind and pressure data from the InSight lander's met station to detect passage of convective vortices and to estimate their diameters. (In spite of considerable effort, InSight did not image any active dust devils.) Figure 8 from that study shows the inferred diameters as a function of time of day. Although there are only a few dozen detections with inferred diameters available, the average diameter does appear to increase after mid-day, with the largest diameters appearing between 14:00 and 15:00 local time. The reason that diameters are not larger earlier in the day may be that boundary layer is not as active as later in day, and so conditions are not as conducive to vortex formation earlier in the day. This result certainly suffers from small number statistics, but field and remote-sensing studies could substantially bolster this dataset simply by reporting the time of day when a dust devil diameter is estimated. \added{Alternative and entirely plausible interpretations are that the deepening of the boundary layer \citep{2015Icar..260..246F,2021AdSpR..67.2219L} or growth of the Monin-Obukhov length \citep{1990JApMe..29..498H, 2005QJRMS.131.1271K} throughout the day allow larger and larger dust devils to occur}, and of course, these effects may operate simultaneously.

Our work here highlights the power of dust devils to inform about conditions in the planetary boundary layer. As discussed in \citet{2015Icar..260..246F}, dust devils can serve as probes of the boundary layer, providing otherwise inaccessible information on this key component of the atmosphere but without requiring in-situ measurements. Imaging from space could suffice. The quality of our inferences, however, rely on the quality of our understanding of the relationships between dust devil properties and ambient conditions. Fortunately, as we point out here, small adjustments and additions to surveying work could provide substantial leverage of these currently obscure connections. Ironically, these opaque phenomena may provide one of the best ways to clarify the behavior of the martian atmosphere.

%% Please use the acknowledgment and contribution environments. This will 
%% be anonomyized when the "anonymous" style option is used. 
\begin{acknowledgments}
BJ was supported by a grant from NASA's Mars Data Analysis Program, solicitation number NNH22ZDA001N-MDAP. CS was supported by a grant from the Idaho Space Grant Consortium. RL acknowledges the support of NASA InSight Participating Scientist Program grant 80NSSC18K1626 and JPL Mars 2020 SuperCam contract 1655893. CZ acknowledges support from the Idaho Space Grant Consortium.

We also acknowledge very helpful input from two anonymous referees.

\end{acknowledgments}

\begin{contribution}
%%This section gives authors the space to recognize author contributions. The text inside this environment is NOT counted towards the total word quanta. At a minimum, manuscripts are expected to include this text:

BJ devised and led the study. RL contributed to analysis. BJ, RL, and LF all contributed to writing this manuscript. All authors contributed to planning and implementing the field work. 

%% But authors are expected to provide more specific details, e.g. 
%%
%%SC was responsible for writing and submitting the manuscript.
%%WWM came up with the initial research concept and edited the manuscript.
%%OTS obtained the funding and edited the manuscript.
%%EBF provided the formal analysis and validation. He also edited the manuscript.
%%GEH Supervised the undergraduates, wrote the software and administers the project github and Zenodo repositories.
%%
%% Authors can use the Contributor Role Taxonomy (CRediT) at
%% for ideas on how write a good statement tailored to their needs.

\end{contribution}


\software{astropy \citep{2013A&A...558A..33A, 2018AJ....156..123A}, matplotlib \citep{Hunter:2007}, numpy \citep{harris2020array}, scipy \citep{2020SciPy-NMeth}}


\bibliography{sample7}{}

\begin{thebibliography}{}
\expandafter\ifx\csname natexlab\endcsname\relax\def\natexlab#1{#1}\fi
\providecommand{\url}[1]{\href{#1}{#1}}
\providecommand{\dodoi}[1]{doi:~\href{http://doi.org/#1}{\nolinkurl{#1}}}
\providecommand{\doeprint}[1]{\href{http://ascl.net/#1}{\nolinkurl{http://ascl.net/#1}}}
\providecommand{\doarXiv}[1]{\href{https://arxiv.org/abs/#1}{\nolinkurl{https://arxiv.org/abs/#1}}}

\bibitem[{S.~P. {Arya}(2001){Arya}}]{2001imm..book.....A}
{Arya}, S.~P. 2001, {Introduction to Micrometeorology}

\bibitem[{ {Astropy Collaboration} {et~al.}(2013){Astropy Collaboration},
  {Robitaille}, {Tollerud}, {Greenfield}, {Droettboom}, {Bray}, {Aldcroft},
  {Davis}, {Ginsburg}, {Price-Whelan}, {Kerzendorf}, {Conley}, {Crighton},
  {Barbary}, {Muna}, {Ferguson}, {Grollier}, {Parikh}, {Nair}, {Unther},
  {Deil}, {Woillez}, {Conseil}, {Kramer}, {Turner}, {Singer}, {Fox}, {Weaver},
  {Zabalza}, {Edwards}, {Azalee Bostroem}, {Burke}, {Casey}, {Crawford},
  {Dencheva}, {Ely}, {Jenness}, {Labrie}, {Lim}, {Pierfederici}, {Pontzen},
  {Ptak}, {Refsdal}, {Servillat}, \& {Streicher}}]{2013A&A...558A..33A}
{Astropy Collaboration}, {Robitaille}, T.~P., {Tollerud}, E.~J., {et~al.} 2013,
  \bibinfo{title}{{Astropy: A community Python package for astronomy},} \aap,
  558, A33, \dodoi{10.1051/0004-6361/201322068}

\bibitem[{ {Astropy Collaboration} {et~al.}(2018){Astropy Collaboration},
  {Price-Whelan}, {Sip{\H{o}}cz}, {G{\"u}nther}, {Lim}, {Crawford}, {Conseil},
  {Shupe}, {Craig}, {Dencheva}, {Ginsburg}, {VanderPlas}, {Bradley},
  {P{\'e}rez-Su{\'a}rez}, {de Val-Borro}, {Aldcroft}, {Cruz}, {Robitaille},
  {Tollerud}, {Ardelean}, {Babej}, {Bach}, {Bachetti}, {Bakanov}, {Bamford},
  {Barentsen}, {Barmby}, {Baumbach}, {Berry}, {Biscani}, {Boquien}, {Bostroem},
  {Bouma}, {Brammer}, {Bray}, {Breytenbach}, {Buddelmeijer}, {Burke},
  {Calderone}, {Cano Rodr{\'\i}guez}, {Cara}, {Cardoso}, {Cheedella}, {Copin},
  {Corrales}, {Crichton}, {D'Avella}, {Deil}, {Depagne}, {Dietrich}, {Donath},
  {Droettboom}, {Earl}, {Erben}, {Fabbro}, {Ferreira}, {Finethy}, {Fox},
  {Garrison}, {Gibbons}, {Goldstein}, {Gommers}, {Greco}, {Greenfield},
  {Groener}, {Grollier}, {Hagen}, {Hirst}, {Homeier}, {Horton}, {Hosseinzadeh},
  {Hu}, {Hunkeler}, {Ivezi{\'c}}, {Jain}, {Jenness}, {Kanarek}, {Kendrew},
  {Kern}, {Kerzendorf}, {Khvalko}, {King}, {Kirkby}, {Kulkarni}, {Kumar},
  {Lee}, {Lenz}, {Littlefair}, {Ma}, {Macleod}, {Mastropietro}, {McCully},
  {Montagnac}, {Morris}, {Mueller}, {Mumford}, {Muna}, {Murphy}, {Nelson},
  {Nguyen}, {Ninan}, {N{\"o}the}, {Ogaz}, {Oh}, {Parejko}, {Parley}, {Pascual},
  {Patil}, {Patil}, {Plunkett}, {Prochaska}, {Rastogi}, {Reddy Janga},
  {Sabater}, {Sakurikar}, {Seifert}, {Sherbert}, {Sherwood-Taylor}, {Shih},
  {Sick}, {Silbiger}, {Singanamalla}, {Singer}, {Sladen}, {Sooley},
  {Sornarajah}, {Streicher}, {Teuben}, {Thomas}, {Tremblay}, {Turner},
  {Terr{\'o}n}, {van Kerkwijk}, {de la Vega}, {Watkins}, {Weaver}, {Whitmore},
  {Woillez}, {Zabalza}, \& {Astropy Contributors}}]{2018AJ....156..123A}
{Astropy Collaboration}, {Price-Whelan}, A.~M., {Sip{\H{o}}cz}, B.~M., {et~al.}
  2018, \bibinfo{title}{{The Astropy Project: Building an Open-science Project
  and Status of the v2.0 Core Package},} \aj, 156, 123,
  \dodoi{10.3847/1538-3881/aabc4f}

\bibitem[{M.~R. {Balme} {et~al.}(2012){Balme}, {Pathare}, {Metzger}, {Towner},
  {Lewis}, {Spiga}, {Fenton}, {Renno}, {Elliott}, {Saca}, {Michaels},
  {Russell}, \& {Verdasca}}]{2012Icar..221..632B}
{Balme}, M.~R., {Pathare}, A., {Metzger}, S.~M., {et~al.} 2012,
  \bibinfo{title}{{Field measurements of horizontal forward motion velocities
  of terrestrial dust devils: Towards a proxy for ambient winds on Mars and
  Earth},} \icarus, 221, 632, \dodoi{10.1016/j.icarus.2012.08.021}

\bibitem[{S.~J. {Conway} {et~al.}(2025){Conway}, {Bickel}, {Fenton}, {Patel},
  {Carson}, {Blouin}, {Crevier}, {Blanc}, {Nguyen}, {Holmes}, {Jackson}, \&
  {Roelofs}}]{2025P&SS..25906072C}
{Conway}, S.~J., {Bickel}, V.~T., {Fenton}, L.~K., {et~al.} 2025,
  \bibinfo{title}{{A global survey for dust devil vortices on mars using MRO
  context camera images enabled by neural networks},} \planss, 259, 106072,
  \dodoi{10.1016/j.pss.2025.106072}

\bibitem[{M.~D. {Ellehoj} {et~al.}(2010){Ellehoj}, {Gunnlaugsson}, {Taylor},
  {Kahanp{\"a}{\"a}}, {Bean}, {Cantor}, {Gheynani}, {Drube}, {Fisher}, {Harri},
  {Holstein-Rathlou}, {Lemmon}, {Madsen}, {Malin}, {Polkko}, {Smith},
  {Tamppari}, {Weng}, \& {Whiteway}}]{2010JGRE..115.0E16E}
{Ellehoj}, M.~D., {Gunnlaugsson}, H.~P., {Taylor}, P.~A., {et~al.} 2010,
  \bibinfo{title}{{Convective vortices and dust devils at the Phoenix Mars
  mission landing site},} Journal of Geophysical Research (Planets), 115,
  E00E16, \dodoi{10.1029/2009JE003413}

\bibitem[{E.~D. {Feigelson} \& G.~J. {Babu}(2012){Feigelson} \&
  {Babu}}]{2012msma.book.....F}
{Feigelson}, E.~D., \& {Babu}, G.~J. 2012, {Modern Statistical Methods for
  Astronomy}, \dodoi{10.48550/arXiv.1205.2064}

\bibitem[{L. {Fenton} {et~al.}(2016){Fenton}, {Reiss}, {Lemmon}, {Marticorena},
  {Lewis}, \& {Cantor}}]{2016SSRv..203...89F}
{Fenton}, L., {Reiss}, D., {Lemmon}, M., {et~al.} 2016,
  \bibinfo{title}{{Orbital Observations of Dust Lofted by Daytime Convective
  Turbulence},} \ssr, 203, 89, \dodoi{10.1007/s11214-016-0243-6}

\bibitem[{L.~K. {Fenton} \& R. {Lorenz}(2015){Fenton} \&
  {Lorenz}}]{2015Icar..260..246F}
{Fenton}, L.~K., \& {Lorenz}, R. 2015, \bibinfo{title}{{Dust devil height and
  spacing with relation to the martian planetary boundary layer thickness},}
  \icarus, 260, 246, \dodoi{10.1016/j.icarus.2015.07.028}

\bibitem[{L.~K. {Fenton} {et~al.}(2022){Fenton}, {Metzger}, {Michaels},
  {Scheidt}, {Dorn}, {Neakrase}, {Cole}, \& {Sprau}}]{2022AeoRe..5900831F}
{Fenton}, L.~K., {Metzger}, S.~M., {Michaels}, T.~I., {et~al.} 2022,
  \bibinfo{title}{{Meteorological and geological controls on dust devil
  activity: Initial results from a field study at Smith Creek Valley, Nevada,
  USA},} Aeolian Research, 59, 100831, \dodoi{10.1016/j.aeolia.2022.100831}

\bibitem[{D. Freedman \& P. Diaconis(1981)Freedman \& Diaconis}]{Freedman1981}
Freedman, D., \& Diaconis, P. 1981, \bibinfo{title}{On the histogram as a
  density estimator:L2 theory,} Zeitschrift f{\"u}r Wahrscheinlichkeitstheorie
  und Verwandte Gebiete, 57, 453, \dodoi{10.1007/BF01025868}

\bibitem[{M. {Golombek} {et~al.}(2023){Golombek}, {Charalambous}, {Newman},
  {Williams}, {Baker}, {Lorenz}, \& {Banerdt}}]{2023LPICo2806.2467G}
{Golombek}, M., {Charalambous}, C., {Newman}, C., {et~al.} 2023, in LPI
  Contributions, Vol. 2806, 54th Lunar and Planetary Science Conference, 2467

\bibitem[{R. {Greeley} {et~al.}(2006){Greeley}, {Whelley}, {Arvidson},
  {Cabrol}, {Foley}, {Franklin}, {Geissler}, {Golombek}, {Kuzmin}, {Landis},
  {Lemmon}, {Neakrase}, {Squyres}, \& {Thompson}}]{2006JGRE..11112S09G}
{Greeley}, R., {Whelley}, P.~L., {Arvidson}, R.~E., {et~al.} 2006,
  \bibinfo{title}{{Active dust devils in Gusev crater, Mars: Observations from
  the Mars Exploration Rover Spirit},} Journal of Geophysical Research
  (Planets), 111, E12S09, \dodoi{10.1029/2006JE002743}

\bibitem[{R. {Greeley} {et~al.}(2010){Greeley}, {Waller}, {Cabrol}, {Landis},
  {Lemmon}, {Neakrase}, {Pendleton Hoffer}, {Thompson}, \&
  {Whelley}}]{2010JGRE..115.0F02G}
{Greeley}, R., {Waller}, D.~A., {Cabrol}, N.~A., {et~al.} 2010,
  \bibinfo{title}{{Gusev Crater, Mars: Observations of three dust devil
  seasons},} Journal of Geophysical Research (Planets), 115, E00F02,
  \dodoi{10.1029/2010JE003608}

\bibitem[{C.~R. Harris {et~al.}(2020)Harris, Millman, van~der Walt, Gommers,
  Virtanen, Cournapeau, Wieser, Taylor, Berg, Smith, Kern, Picus, Hoyer, van
  Kerkwijk, Brett, Haldane, del R{\'{i}}o, Wiebe, Peterson,
  G{\'{e}}rard-Marchant, Sheppard, Reddy, Weckesser, Abbasi, Gohlke, \&
  Oliphant}]{harris2020array}
Harris, C.~R., Millman, K.~J., van~der Walt, S.~J., {et~al.} 2020,
  \bibinfo{title}{Array programming with {NumPy},} Nature, 585, 357,
  \dodoi{10.1038/s41586-020-2649-2}

\bibitem[{G.~D. {Hess} \& K.~T. {Spillane}(1990){Hess} \&
  {Spillane}}]{1990JApMe..29..498H}
{Hess}, G.~D., \& {Spillane}, K.~T. 1990, \bibinfo{title}{{Characteristics of
  Dust Devils in Australia.},} Journal of Applied Meteorology, 29, 498,
  \dodoi{10.1175/1520-0450(1990)029<0498:CODDIA>2.0.CO;2}

\bibitem[{J.~D. Hunter(2007)Hunter}]{Hunter:2007}
Hunter, J.~D. 2007, \bibinfo{title}{Matplotlib: A 2D graphics environment,}
  Computing In Science \& Engineering, 9, 90

\bibitem[{B. {Jackson}(2020){Jackson}}]{2020Icar..33813523J}
{Jackson}, B. 2020, \bibinfo{title}{{On the relationship between dust devil
  radii and heights},} \icarus, 338, 113523,
  \dodoi{10.1016/j.icarus.2019.113523}

\bibitem[{B. {Jackson} {et~al.}(2021){Jackson}, {Crevier}, {Szurgot}, {Battin},
  {Perrin}, \& {Rodriguez}}]{2021PSJ.....2..206J}
{Jackson}, B., {Crevier}, J., {Szurgot}, M., {et~al.} 2021,
  \bibinfo{title}{{Inferring Vortex and Dust Devil Statistics from InSight},}
  \psj, 2, 206, \dodoi{10.3847/PSJ/ac260d}

\bibitem[{B. {Jackson} {et~al.}(2018){Jackson}, {Lorenz}, {Davis}, \&
  {Lipple}}]{2018RemS...10...65J}
{Jackson}, B., {Lorenz}, R., {Davis}, K., \& {Lipple}, B. 2018,
  \bibinfo{title}{{Using an Instrumented Drone to Probe Dust Devils on Oregon's
  Alvord Desert.},} Remote Sensing, 10, 65, \dodoi{10.3390/rs10010065}

\bibitem[{B. {Jackson} {et~al.}(2025){Jackson}, {Szurgot}, {Fenton}, {Lorenz},
  {Gambill}, \& {Arzaga}}]{2025LPSC}
{Jackson}, B., {Szurgot}, C., {Fenton}, L., {et~al.} 2025, in LPI
  Contributions, LPI Contributions

\bibitem[{M.~A. {Kahre} {et~al.}(2017){Kahre}, {Murphy}, {Newman}, {Wilson},
  {Cantor}, {Lemmon}, \& {Wolff}}]{2017acm..book..229K}
{Kahre}, M.~A., {Murphy}, J.~R., {Newman}, C.~E., {et~al.} 2017, in Asteroids,
  Comets, Meteors - ACM2017, ed. R.~M. {Haberle}, R.~T. {Clancy}, F.~{Forget},
  M.~D. {Smith}, \& R.~W. {Zurek}, 229--294, \dodoi{10.1017/9781139060172.010}

\bibitem[{K.~M. {Kanak}(2005){Kanak}}]{2005QJRMS.131.1271K}
{Kanak}, K.~M. 2005, \bibinfo{title}{{Numerical simulation of dust devil-scale
  vortices},} Quarterly Journal of the Royal Meteorological Society, 131, 1271,
  \dodoi{10.1256/qj.03.172}

\bibitem[{A. {Kleinb{\"o}hl} {et~al.}(2024){Kleinb{\"o}hl}, {Willacy},
  {Slipski}, {Poncin}, {Halekas}, \& {Mayyasi}}]{2024NatAs...8..827K}
{Kleinb{\"o}hl}, A., {Willacy}, K., {Slipski}, M.~J., {et~al.} 2024,
  \bibinfo{title}{{Hydrogen escape on Mars dominated by water vapour photolysis
  above the hygropause},} Nature Astronomy, 8, 827,
  \dodoi{10.1038/s41550-024-02268-x}

\bibitem[{K.~H. {Knuth}(2006){Knuth}}]{2006physics...5197K}
{Knuth}, K.~H. 2006, \bibinfo{title}{{Optimal Data-Based Binning for
  Histograms},} arXiv e-prints, physics/0605197,
  \dodoi{10.48550/arXiv.physics/0605197}

\bibitem[{M.~V. {Kurgansky}(2006){Kurgansky}}]{2006GeoRL..3319S06K}
{Kurgansky}, M.~V. 2006, \bibinfo{title}{{Steady-state properties and
  statistical distribution of atmospheric dust devils},} \grl, 33, L19S06,
  \dodoi{10.1029/2006GL026142}

\bibitem[{M.~V. {Kurgansky}(2012){Kurgansky}}]{2012Icar..219..556K}
{Kurgansky}, M.~V. 2012, \bibinfo{title}{{Statistical distribution of
  atmospheric dust devils},} \icarus, 219, 556,
  \dodoi{10.1016/j.icarus.2012.04.006}

\bibitem[{M.~V. {Kurgansky}(2018){Kurgansky}}]{2018Icar..300...97K}
{Kurgansky}, M.~V. 2018, \bibinfo{title}{{To the theory of particle lifting by
  terrestrial and Martian dust devils},} \icarus, 300, 97,
  \dodoi{10.1016/j.icarus.2017.08.029}

\bibitem[{M.~V. {Kurgansky}(2019){Kurgansky}}]{2019Icar..317..209K}
{Kurgansky}, M.~V. 2019, \bibinfo{title}{{On the statistical distribution of
  pressure drops in convective vortices: Applications to Martian dust devils},}
  \icarus, 317, 209, \dodoi{10.1016/j.icarus.2018.08.004}

\bibitem[{M.~V. {Kurgansky} {et~al.}(2016){Kurgansky}, {Lorenz}, {Renno},
  {Takemi}, {Gu}, \& {Wei}}]{2016SSRv..203..209K}
{Kurgansky}, M.~V., {Lorenz}, R.~D., {Renno}, N.~O., {et~al.} 2016,
  \bibinfo{title}{{Dust Devil Steady-State Structure from a Fluid Dynamics
  Perspective},} \ssr, 203, 209, \dodoi{10.1007/s11214-016-0281-0}

\bibitem[{R. {Lorenz}(2011){Lorenz}}]{2011Icar..215..381L}
{Lorenz}, R. 2011, \bibinfo{title}{{On the statistical distribution of dust
  devil diameters},} \icarus, 215, 381, \dodoi{10.1016/j.icarus.2011.06.005}

\bibitem[{R. {Lorenz}(2013){Lorenz}}]{2013Icar..226..964L}
{Lorenz}, R. 2013, \bibinfo{title}{{The longevity and aspect ratio of dust
  devils: Effects on detection efficiencies and comparison of landed and
  orbital imaging at Mars},} \icarus, 226, 964,
  \dodoi{10.1016/j.icarus.2013.06.031}

\bibitem[{R.~D. {Lorenz}(2009){Lorenz}}]{2009Icar..203..683L}
{Lorenz}, R.~D. 2009, \bibinfo{title}{{Power law of dust devil diameters on
  Mars and Earth},} \icarus, 203, 683, \dodoi{10.1016/j.icarus.2009.06.029}

\bibitem[{R.~D. {Lorenz}(2012){Lorenz}}]{2012GI......1..209L}
{Lorenz}, R.~D. 2012, \bibinfo{title}{{Observing desert dust devils with a
  pressure logger},} Geoscientific Instrumentation, Methods and Data Systems,
  1, 209, \dodoi{10.5194/gi-1-209-201210.5194/gid-2-477-2012}

\bibitem[{R.~D. {Lorenz}(2016){Lorenz}}]{2016Icar..271..326L}
{Lorenz}, R.~D. 2016, \bibinfo{title}{{Heuristic estimation of dust devil
  vortex parameters and trajectories from single-station meteorological
  observations: Application to InSight at Mars},} \icarus, 271, 326,
  \dodoi{10.1016/j.icarus.2016.02.001}

\bibitem[{R.~D. {Lorenz}(2021){Lorenz}}]{2021AdSpR..67.2219L}
{Lorenz}, R.~D. 2021, \bibinfo{title}{{An engineering model of Titan surface
  winds for Dragonfly landed operations},} Advances in Space Research, 67,
  2219, \dodoi{10.1016/j.asr.2021.01.023}

\bibitem[{R.~D. {Lorenz} \& B.~K. {Jackson}(2016){Lorenz} \&
  {Jackson}}]{2016SSRv..203..277L}
{Lorenz}, R.~D., \& {Jackson}, B.~K. 2016, \bibinfo{title}{{Dust Devil
  Populations and Statistics},} \ssr, 203, 277,
  \dodoi{10.1007/s11214-016-0277-9}

\bibitem[{L.~D.~V. {Neakrase} \& R. {Greeley}(2010){Neakrase} \&
  {Greeley}}]{2010Icar..206..306N}
{Neakrase}, L. D.~V., \& {Greeley}, R. 2010, \bibinfo{title}{{Dust devil
  sediment flux on Earth and Mars: Laboratory simulations},} \icarus, 206, 306,
  \dodoi{10.1016/j.icarus.2009.08.028}

\bibitem[{C.~E. {Newman} {et~al.}(2019){Newman}, {Kahanp{\"a}{\"a}},
  {Richardson}, {Mart{\'\i}nez}, {Vicente-Retortillo}, \&
  {Lemmon}}]{2019JGRE..124.3442N}
{Newman}, C.~E., {Kahanp{\"a}{\"a}}, H., {Richardson}, M.~I., {et~al.} 2019,
  \bibinfo{title}{{MarsWRF Convective Vortex and Dust Devil Predictions for
  Gale Crater Over 3 Mars Years and Comparison With MSL-REMS Observations},}
  Journal of Geophysical Research (Planets), 124, 3442,
  \dodoi{10.1029/2019JE006082}

\bibitem[{C.~E. {Newman} \& M.~I. {Richardson}(2015){Newman} \&
  {Richardson}}]{2015Icar..257...47N}
{Newman}, C.~E., \& {Richardson}, M.~I. 2015, \bibinfo{title}{{The impact of
  surface dust source exhaustion on the martian dust cycle, dust storms and
  interannual variability, as simulated by the MarsWRF General Circulation
  Model},} \icarus, 257, 47, \dodoi{10.1016/j.icarus.2015.03.030}

\bibitem[{C.~E. Newman {et~al.}(2022)Newman, Bertrand, Fenton, Guzewich,
  Jackson, Lewis, Mischna, Montabone, \& Wellington}]{NEWMAN2022637}
Newman, C.~E., Bertrand, T., Fenton, L.~K., {et~al.} 2022, in Treatise on
  Geomorphology (Second Edition), second edition edn., ed. J.~J.~F. Shroder
  (Oxford: Academic Press), 637--666,
  \dodoi{https://doi.org/10.1016/B978-0-12-818234-5.00143-7}

\bibitem[{A.~V. {Pathare} {et~al.}(2010){Pathare}, {Balme}, {Metzger}, {Spiga},
  {Towner}, {Renno}, \& {Saca}}]{2010Icar..209..851P}
{Pathare}, A.~V., {Balme}, M.~R., {Metzger}, S.~M., {et~al.} 2010,
  \bibinfo{title}{{Assessing the power law hypothesis for the size-frequency
  distribution of terrestrial and martian dust devils},} \icarus, 209, 851,
  \dodoi{10.1016/j.icarus.2010.06.027}

\bibitem[{S. {Rafkin} {et~al.}(2016){Rafkin}, {Jemmett-Smith}, {Fenton},
  {Lorenz}, {Takemi}, {Ito}, \& {Tyler}}]{2016SSRv..203..183R}
{Rafkin}, S., {Jemmett-Smith}, B., {Fenton}, L., {et~al.} 2016,
  \bibinfo{title}{{Dust Devil Formation},} \ssr, 203, 183,
  \dodoi{10.1007/s11214-016-0307-7}

\bibitem[{N.~O. {Renn{\'o}} {et~al.}(1998){Renn{\'o}}, {Burkett}, \&
  {Larkin}}]{1998JAtS...55.3244R}
{Renn{\'o}}, N.~O., {Burkett}, M.~L., \& {Larkin}, M.~P. 1998,
  \bibinfo{title}{{A Simple Thermodynamical Theory for Dust Devils.},} Journal
  of the Atmospheric Sciences, 55, 3244,
  \dodoi{10.1175/1520-0469(1998)055<3244:ASTTFD>2.0.CO;2}

\bibitem[{N.~O. {Renn{\'o}} \& A.~P. {Ingersoll}(1996){Renn{\'o}} \&
  {Ingersoll}}]{1996JAtS...53..572R}
{Renn{\'o}}, N.~O., \& {Ingersoll}, A.~P. 1996, \bibinfo{title}{{Natural
  Convection as a Heat Engine: A Theory for CAPE.},} Journal of the Atmospheric
  Sciences, 53, 572, \dodoi{10.1175/1520-0469(1996)053<0572:NCAAHE>2.0.CO;2}

\bibitem[{M.~I. {Richardson} {et~al.}(2007){Richardson}, {Toigo}, \&
  {Newman}}]{2007JGRE..112.9001R}
{Richardson}, M.~I., {Toigo}, A.~D., \& {Newman}, C.~E. 2007,
  \bibinfo{title}{{PlanetWRF: A general purpose, local to global numerical
  model for planetary atmospheric and climate dynamics},} Journal of
  Geophysical Research (Planets), 112, E09001, \dodoi{10.1029/2006JE002825}

\bibitem[{J.~A. {Ryan} \& J.~J. {Carroll}(1970){Ryan} \&
  {Carroll}}]{1970JGR....75..531R}
{Ryan}, J.~A., \& {Carroll}, J.~J. 1970, \bibinfo{title}{{Dust devil wind
  velocities: Mature state},} \jgr, 75, 531, \dodoi{10.1029/JC075i003p00531}

\bibitem[{J.~D. {Scargle} {et~al.}(2013){Scargle}, {Norris}, {Jackson}, \&
  {Chiang}}]{2013ApJ...764..167S}
{Scargle}, J.~D., {Norris}, J.~P., {Jackson}, B., \& {Chiang}, J. 2013,
  \bibinfo{title}{{Studies in Astronomical Time Series Analysis. VI. Bayesian
  Block Representations},} \apj, 764, 167, \dodoi{10.1088/0004-637X/764/2/167}

\bibitem[{D.~W. Scott(1979)Scott}]{ScottsRule}
Scott, D.~W. 1979, \bibinfo{title}{On optimal and data-based histograms,}
  Biometrika, 66, 605, \dodoi{10.1093/biomet/66.3.605}

\bibitem[{P.~C. {Sinclair}(1966){Sinclair}}]{1966PhDT........31S}
{Sinclair}, P.~C. 1966, PhD thesis, University of Arizona

\bibitem[{C. {Stanzel} {et~al.}(2008){Stanzel}, {P{\"a}tzold}, {Williams},
  {Whelley}, {Greeley}, {Neukum}, \& {HRSC Co-Investigator
  Team}}]{2008Icar..197...39S}
{Stanzel}, C., {P{\"a}tzold}, M., {Williams}, D.~A., {et~al.} 2008,
  \bibinfo{title}{{Dust devil speeds, directions of motion and general
  characteristics observed by the Mars Express High Resolution Stereo Camera},}
  \icarus, 197, 39, \dodoi{10.1016/j.icarus.2008.04.017}

\bibitem[{A.~D. {Toigo} {et~al.}(2012){Toigo}, {Lee}, {Newman}, \&
  {Richardson}}]{2012Icar..221..276T}
{Toigo}, A.~D., {Lee}, C., {Newman}, C.~E., \& {Richardson}, M.~I. 2012,
  \bibinfo{title}{{The impact of resolution on the dynamics of the martian
  global atmosphere: Varying resolution studies with the MarsWRF GCM},}
  \icarus, 221, 276, \dodoi{10.1016/j.icarus.2012.07.020}

\bibitem[{P. Virtanen {et~al.}(2020)Virtanen, Gommers, Oliphant, Haberland,
  Reddy, Cournapeau, Burovski, Peterson, Weckesser, Bright, {van der Walt},
  Brett, Wilson, Millman, Mayorov, Nelson, Jones, Kern, Larson, Carey, Polat,
  Feng, Moore, {VanderPlas}, Laxalde, Perktold, Cimrman, Henriksen, Quintero,
  Harris, Archibald, Ribeiro, Pedregosa, {van Mulbregt}, \& {SciPy 1.0
  Contributors}}]{2020SciPy-NMeth}
Virtanen, P., Gommers, R., Oliphant, T.~E., {et~al.} 2020,
  \bibinfo{title}{{{SciPy} 1.0: Fundamental Algorithms for Scientific Computing
  in Python},} Nature Methods, 17, 261, \dodoi{10.1038/s41592-019-0686-2}

\bibitem[{G.~E. {Willis} \& J.~W. {Deardorff}(1979){Willis} \&
  {Deardorff}}]{1979JGR....84..295W}
{Willis}, G.~E., \& {Deardorff}, J.~W. 1979, \bibinfo{title}{{Laboratory
  observations of turbulent penetrative-convection planforms},} \jgr, 84, 295,
  \dodoi{10.1029/JC084iC01p00295}

\end{thebibliography}
\bibliographystyle{aasjournalv7}


% \listofchanges

\end{document}